\documentclass[11pt,twoside]{article}

\usepackage{asp2004}
\usepackage{epsf}
\usepackage{lscape}
\usepackage{amsmath}
\usepackage{amssymb}

\newcommand{\mdot}{\ensuremath{\dot{M}}}                             

\begin{document}
\title{The Influence of X-ray Emission on the Stellar Wind of O~Stars}    
\author{Ji\v{r}\'{\i} Krti\v{c}ka}   
\affil{\'Ustav teoretick\'e fyziky a astrofyziky, P\v r\'\i rodov\v edeck\'a\
fakulta Masarykovy univerzity, Kotl\'a\v rsk\'a 2, CZ-611 37 Brno,
Czech Republic}    
\author{Ji\v{r}\'{\i} Kub\'at}
\affil{Astronomick\'y \'ustav, Akademie v\v{e}d
        \v{C}esk\'e republiky, CZ-251 65 Ond\v{r}ejov, Czech Republic}

\begin{abstract} 
We study the influence of X-ray radiation on the wind parameters of O stars. For
this purpose we use our own NLTE wind code. The X-ray emission (assumed to be
generated in wind shocks) is treated as an input quantity. We study its
influence on the mass-loss rate, terminal velocity and ionization state of the
stellar wind of Galactic O stars.
\end{abstract}

\section{Introduction}
From the observation of hot stars it is known that they emit X-ray radiation
\citet[e.g. Bergh\"ofer et al.,] [hereafter BSC] {rosat}. O star X-ray
luminosity $L_\text{X}$ scales with stellar luminosity $L$ roughly as
$L_\text{X}\approx10^{-7}L$.
The X-rays in O~stars are generated
most likely by shocks that emerge in the supersonic stellar wind either due to
the wind instability \citep{felpulpal} or due to the stellar magnetic field
\citep{udo}.

\section{Wind models}

NLTE stationary spherically symmetric wind models applied here were described by
\cite{nltei}. To enable reliable calculation of wind models in the case when
strong source of X-ray radiation is present, we included also the Auger
ionization into NLTE equations and we extended the set of model ionization
stages.

We assume that a part of wind material is heated by the shock to a very
high temperature 
of the order of $10^6\,$K.
The shock temperature is given by the Rankine-Hugoniot shock condition.
We assume that shock velocity difference is proportional to the wind velocity
with the multiplicative coefficient as a free parameter.
The density of heated material $\rho_\text{x}$ is related to the ambient wind
density $\rho$ by $\rho_\text{x}=f_\text{x}^{1/2}\rho$
where $f_\text{x}$ is the fillig factor (second free parameter). We include the
shock emissivity
into
a
corresponding emission coefficient.

Parameters of studied O stars
were adopted from Repolust et al. (\citeyear{rep}), Markova
et al. (\citeyear{upice}) and Lamers et al. (\citeyear{lsl}). X-ray luminosities
are from 
BSC.


\section{The origin of $L_\text{x}\sim10^{-7}L$ relation}
\begin{figure}[ht]
\plottwo{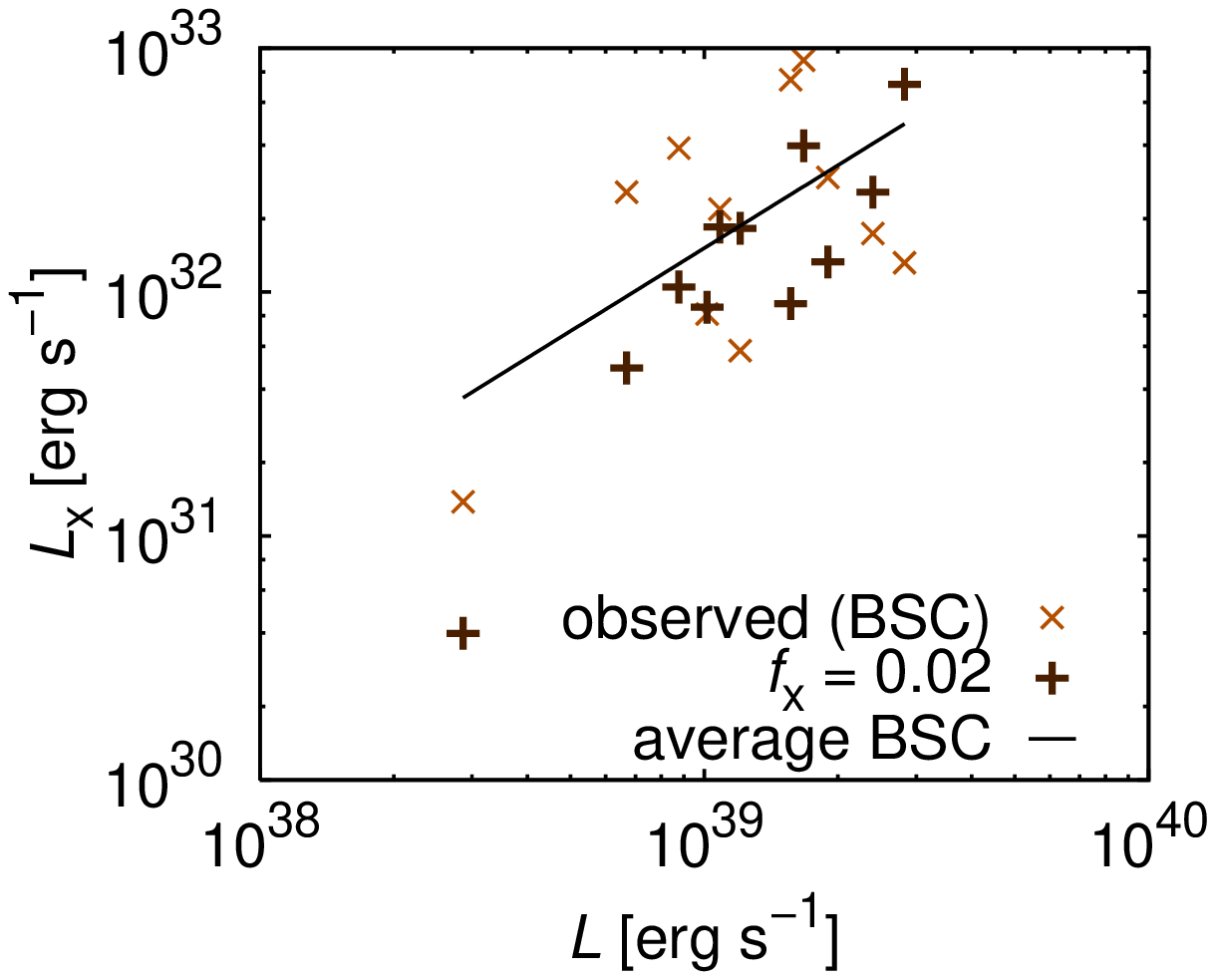}{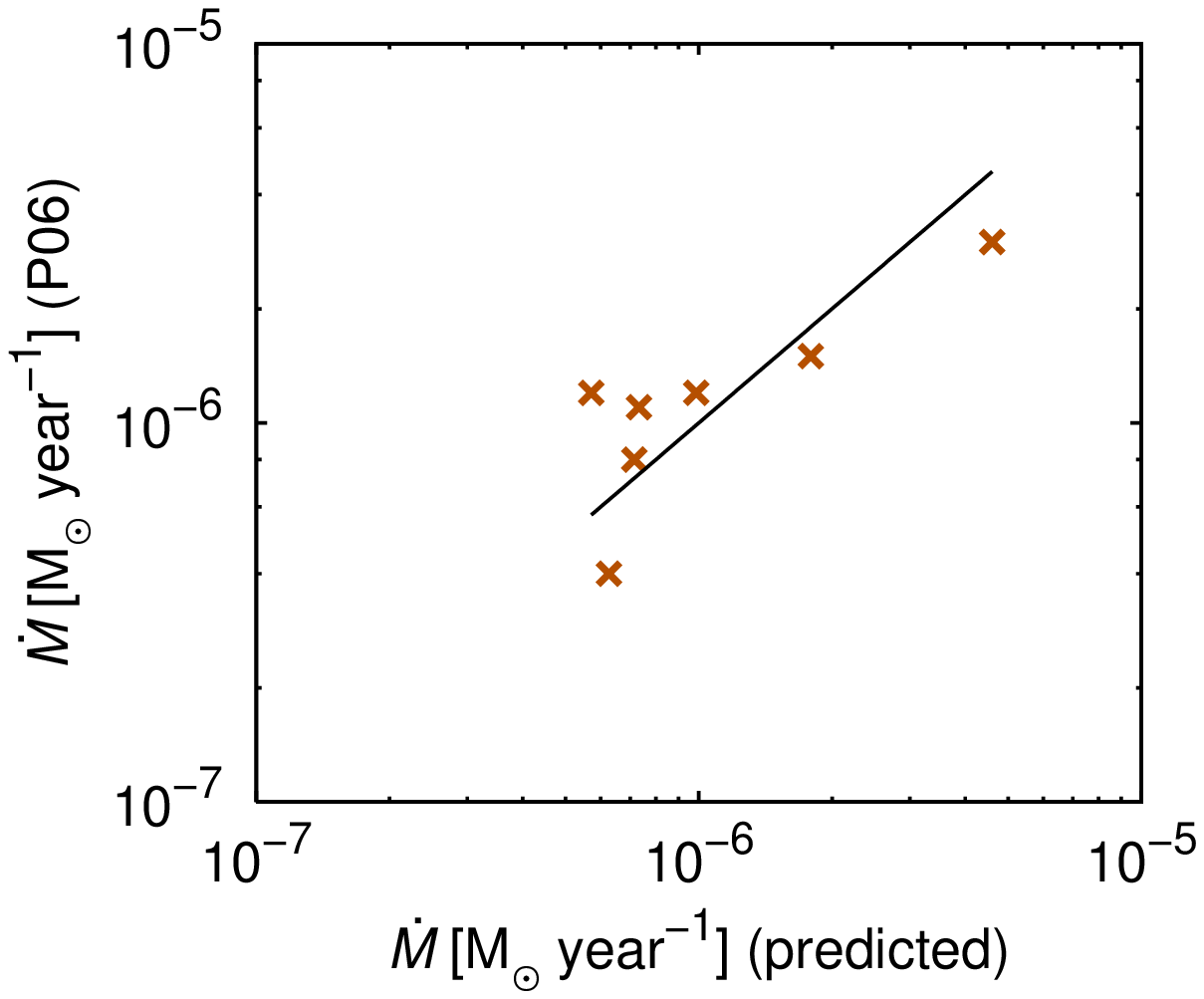}
\caption{Left panel: Comparison of
a
calculated and observed X-ray luminosity
($f_\text{x}=0.02$). Right panel:
Comparison
of
predicted mass-loss rates
\mdot\ and
its
upper limits
derived
from observation
by \citet[P06 in the figure]{pulchuch}.
Note a good agreement between these rates.}
\label{obrazek}
\end{figure}

We calculated models with the same filling factor $f_\text{x}=0.02$
for all stars
(see Fig.~\ref{obrazek}).
For stars with luminosities
$L\gtrsim10^{39}\,\text{erg}\,\text{s}^{-1}$ this
relativelly well resembles the relation between $L_\text{x}$ and $L$ derived by
BSC.

%
%

\section{Influence of X-rays on wind parameters}

Because the mass loss rate is determined at low wind
velocities (where the ionization state is not significantly
influenced by X-rays), the O star mass-loss rate is not
influenced by the presence of X-rays. This may change
for stars with lower luminosities or for stars with weaker ionizing radiation,
i.e.~for B stars.


X-rays are generated in the outer wind regions and do not penetrate deep into
the inner ones. Because in these outer regions the terminal velocity
%
is being determined, the presence of X-rays 
may influence its value
(especially
for stars with lower effective temperature) due to changed ionization
in the outer wind.

\acknowledgements 
Grants GA\,\v{C}R 205/03/D020, 205/04/1267.

%
%
%
%


\end{document}